\documentclass[acus]{JAC2003}

\usepackage{graphicx}
\usepackage{booktabs}

\begin{document}
\title{ EFFECT OF COMPTON SCATTERING ON THE ELECTRON BEAM DYNAMICS AT THE ATF DAMPING RING\thanks{Work supported by Agence nationale de la recherche (ANR)}}

\author{I. Chaikovska\thanks{chaikovs@lal.in2p3.fr}, C. Bruni, N. Delerue, A. Variola, F. Zomer LAL,  University Paris-Sud XI, Orsay, France\\
 K. Kubo, T. Naito, T. Omori, N. Terunuma, J. Urakawa KEK, Ibaraki, Japan}

\maketitle

\begin{abstract}

Compton scattering provides one of the most promising scheme to obtain polarized positrons for the next generation of  $e^-$ -- $e^+$ colliders~\cite{omori}. Moreover it is an attractive method to produce monochromatic high energy polarized gammas for nuclear applications ~\cite{nuclear} and X-rays for compact light sources~\cite{thomX}. In this framework a four-mirror Fabry-P\'erot cavity has been installed at the Accelerator Test Facility (ATF - KEK, Tsukuba, Japan~\cite{ATF}) and is used to produce an intense flux of polarized gamma rays by Compton scattering~\cite{ipac-mightylaser}. For electrons at the ATF energy (1.28 GeV) Compton scattering may result in a shorter lifetime due to the limited bucket acceptance. We have implemented the effect of Compton scattering on a 2D tracking code with a Monte-Carlo method. This code has been used to study the longitudinal dynamics of the electron beam at the ATF damping ring, in particular the evolution of the energy spread and the bunch length under Compton scattering. The results obtained are presented and discussed. Possible methods to observe the effect of Compton scattering on the ATF beam are proposed.

\end{abstract}

\section{Introduction}

For the future linear collider projects (ILC, CLIC), positron sources are considered as a critical and challenging component. This is essentially due to the very high beam intensity required, orders of magnitude higher than existing ones. The main solution to generate polarized positrons is by using high-energy polarized photons (gamma rays). These gamma rays can be produced by Compton scattering and be converted into $e^-$ -- $e^+$ pairs. To meet the requirements imposed by these future projects, one needs to increase significantly the flux of the gamma rays produced. This can be done by using a high average power laser system based on a Fabry-P\'erot cavity and a high current electron beam.  

In this framework, a four-mirror Fabry-P\'erot cavity has been built and successfully installed at the Accelerator Test Facility (ATF) at KEK (Japan) to test the production of gamma rays ~\cite{ipac-mightylaser}.

During the collision between electrons and laser photons, energy is transferred. This results in an increased energy spread for the electron bunch. In a storage ring this increases the energy spread turn by turn and may have significant effect on the electron beam dynamics, reducing the gamma ray flux despite the synchrotron damping.
Therefore, it is important to investigate this effect in the context of both the future positron sources and the accelerator--based compact light sources using Compton scattering~\cite{thomX}. 

\section{Longitudinal beam dynamics with Compton scattering}
\subsection{Longitudinal dynamics in a ring}
During a revolution an electron looses a fraction of its energy by synchrotron radiation. To keep the electron beam in the ring, this energy loss has to be compensated by the energy given by the radio frequency (RF) cavities. The acceleration process by itself results in damping of the longitudinal oscillations in such a way that the trajectory of each electron tends toward center of the RF bucket. The damping is limited due to continuous excitation, so-called quantum excitation. The balance between the radiation damping and  quantum excitation is reached at equilibrium defining an area occupied by the electrons in the RF bucket where the particle oscillations are stable. 

All processes mentioned above, characterize the longitudinal dynamics of a beam in a storage ring like the ATF damping ring (DR). However, the presence of Compton collisions affects this dynamics.

Longitudinal beam dynamics can be described by two coupled variables related to the RF acceleration process and defined with respect to the synchronous particle:
\begin{itemize}
\item energy gained by the electron $j$, $\delta E_j= E_j - E_s$, where $E_j$ and $E_s$ are the energies of any given electron in the bunch and that of the synchronous particle respectively.
\item longitudinal displacement from the synchronous particle that can be expressed also by the time difference $\tau_j = \tau_j - \tau_s$ between the arrival time of the electron at any particular position and the arrival time of the synchronous particle.
\end{itemize}
Since the synchronous particle has zero time difference and has the nominal energy, it is possible to use it as a reference system. 
\subsection{Simulation}
\begin{figure*}[tb]
    \centering
   \includegraphics*[height=9.5cm]{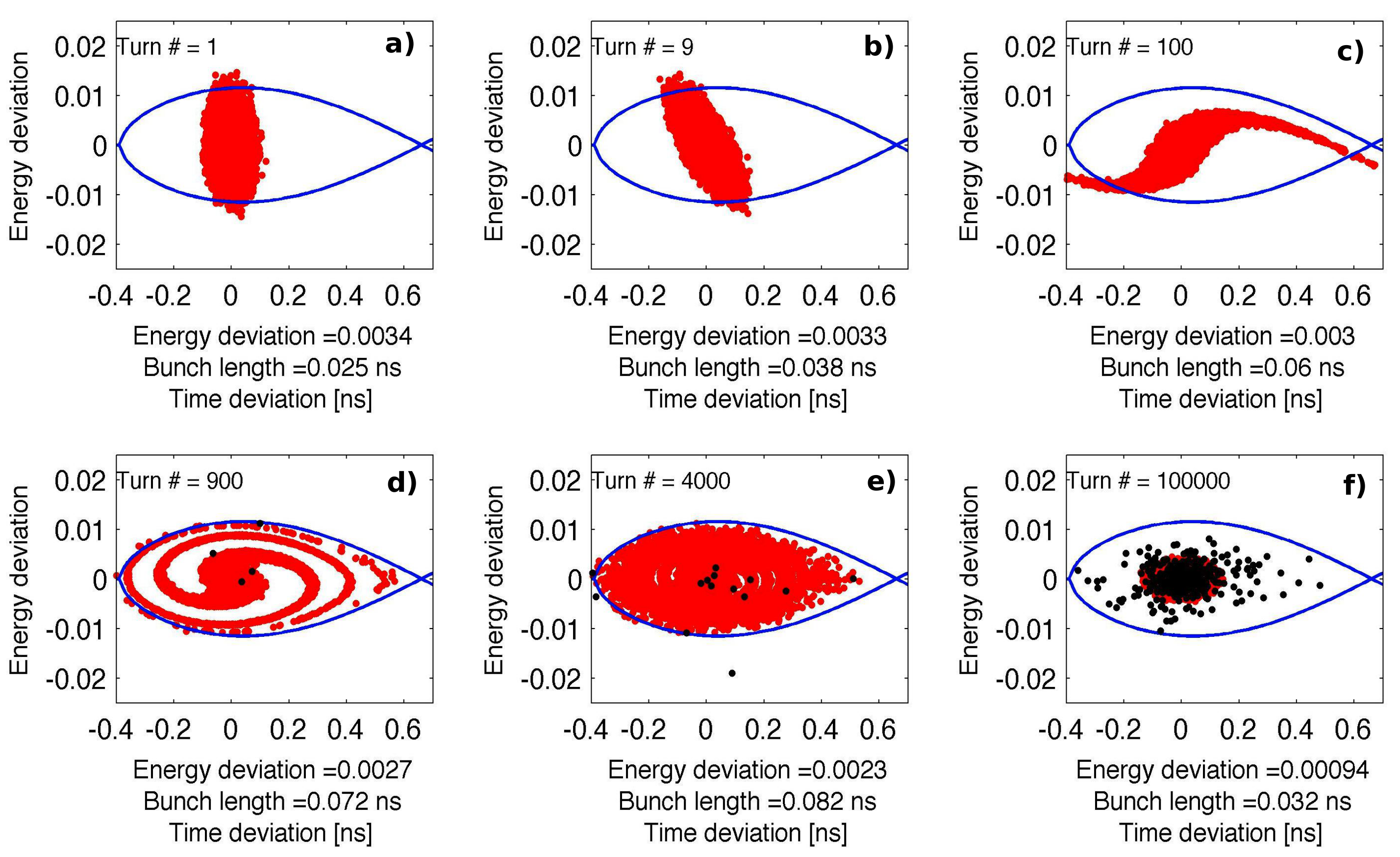}
    \caption{Evolution of the longitudinal phase space of one electron bunch at the ATF DR  with Compton scattering. The bunch is modeled by  $10^5$ macro-particles and the laser power is 100 kW. The blue curve defines the separatrix. Macro-particles denoted in black underwent Compton scattering at least once.}
    \label{fig:phs}
\end{figure*}
The motion of electrons  in the longitudinal phase space is described by a second order non linear differential equation. As a result, electrons oscillate  in longitudinal position and in energy  with respect to the synchronous particle.

In our simulation code we divide  the electron beam in a large number of macro-particles ($10^5$ MP). To find the oscillation amplitudes, we use a numerical solution of the equations of motion in which the energy change and time displacement per revolution are given by
\begin{eqnarray}\label{eq:energy}
&& \delta E_{i,n+1} = \delta E_{i,n} - U_{synchRad} + V_0 sin(\omega_{RF}t + \phi_s)-
\nonumber \\
&&  \qquad \qquad    - \delta E_{damp} - \delta E_{compt} + \delta E_{qe} \nonumber \\
&& \tau_{i,n+1} = \tau_{i,n} - \alpha \frac{\delta E_{i,n}}{E},
\end{eqnarray}
where $ \delta E_{i,n} $ and $ \tau_{i,n}$ are respectively the energy deviation and time displacement of the $i$-th macro-particle on the $n$-th turn, $U_{synchRad}$ is the energy loss per turn due to synchrotron radiation, $ V_0$ is the peak voltage of RF cavity, $\phi_s$ is the synchronous phase,  $ \delta E_{damp} $ is the fraction of energy deviation due to damping, $\delta E_{compt}$ is the fraction of energy deviation due to Compton scattering, $\delta E_{qe}$ is the  fraction of energy deviation given by the quantum excitations, $ \alpha $ is the first order momentum compaction factor and $E$ is the electron energy. Reference~\cite{Sands} contains a full description of each term in Eq.~\ref{eq:energy}. 

The $\delta E_{compt}$ term is calculated for each macro-particle on a turn by turn basis using a Monte-Carlo technique to decide if a given macro-particle has interacted and how much energy was lost due to Compton scattering. A spectrum of scattered gamma rays is obtained by using the simulation code CAIN~\cite{cain}.
\begin{table}[hbt]
\centering
\caption{ATF parameters}
\begin{tabular}{lcc}
\toprule
\textbf{Description} & \textbf{Value} \\ 
Electron energy, $E$         & 1.28\,GeV \\
Electron charge, $C_e$      & 1.6\,nC \\
Electron bunch length at injection, $\tau_{e0}$  & 25\, ps \\
Momentum compaction factor, $\alpha$  & 0.002085 \\
Harmonic number, $h$  & 330 \\
Peak RF Voltage, $V_0$  & 250\,kV  \\
Longitudinal damping time, $t_{damp}$  & 19.5\,ms  \\
Energy spread at injection, $\epsilon_0$ & 0.8\,\% FWHM \\
Revolution period, $T_0$ & 463\, ns \\
Synchronous phase, $\phi_s$ & 10.2\, deg.  \\
Energy loss per turn, $U_{synchRad}$     & 44\, keV \\
Synchrotron frequency, $\Omega_s$  & 10\, kHz  \\
RF bucket height, $(\frac{\Delta E}{E})_{max}$  & 0.012  \\
Laser photon energy, $E_{ph}$  & 1.2\, eV  \\
Crossing angle, $\theta$  & 8\, deg.  \\
Laser pulse length, $\tau_L$  & 20\, ps  \\
 \bottomrule
 \end{tabular}
\label{table:parameters}
\end{table}

Table~\ref{table:parameters} summarises the parameters used for the simulations.
We do not expect other effects to significantly affect	the longitudinal beam dynamics in the ATF DR.
\section{Results and discussion} 

Using our simulation code, we studied the motion of the electrons after injection in the DR. On Fig.~\ref{fig:phs} the turn by turn evolution of the electron bunch injected into such bucket is shown as well as the boundary of the bucket (separatrix). When the energy change for an electron $\delta E$ brings it beyond the separatrix, the electron is ejected from the bunch and eventually lost.

One can see on Fig.~\ref{fig:phs} the  effect of bunch mismatching inside the bucket. The injected bunch is shorter than the RF bucket but has a large energy spread, therefore, under the influence of the RF field it rotates in the longitudinal phase space (see Fig.~\ref{fig:phs}: a, b, c). As the electron bunch is mismatched, it filamentates:  the electrons spread out in the phase space and evolve into a spiral bounded by the separatrix. After  $\sim\!\!1\times10^3$  of turns the bunch fills the RF bucket and the damping process starts. After $\sim\!2\times10^5$~turns (90 ms) the equilibrium between quantum excitation and damping is reached. However, this equilibrium is modified by Compton scattering (see Fig.~\ref{fig:phs}: d, e, f).

In the absence of Compton scattering, at equilibrium all electrons remain within the separatrix. Bucket height is $\sim$15~MeV, so low energy Compton scattering will displace the electrons within the separatrix (see Fig.~\ref{fig:phs} d, the scattered electrons are tagged in black). Whereas a larger perturbations such as high energy Compton Scattering ($\Delta E > 15$ MeV) will kick some electrons out and those outside the bucket will get lost (see  Fig.~\ref{fig:phs} e).

\begin{figure}[htb]
   \centering
   \includegraphics*[width=65mm]{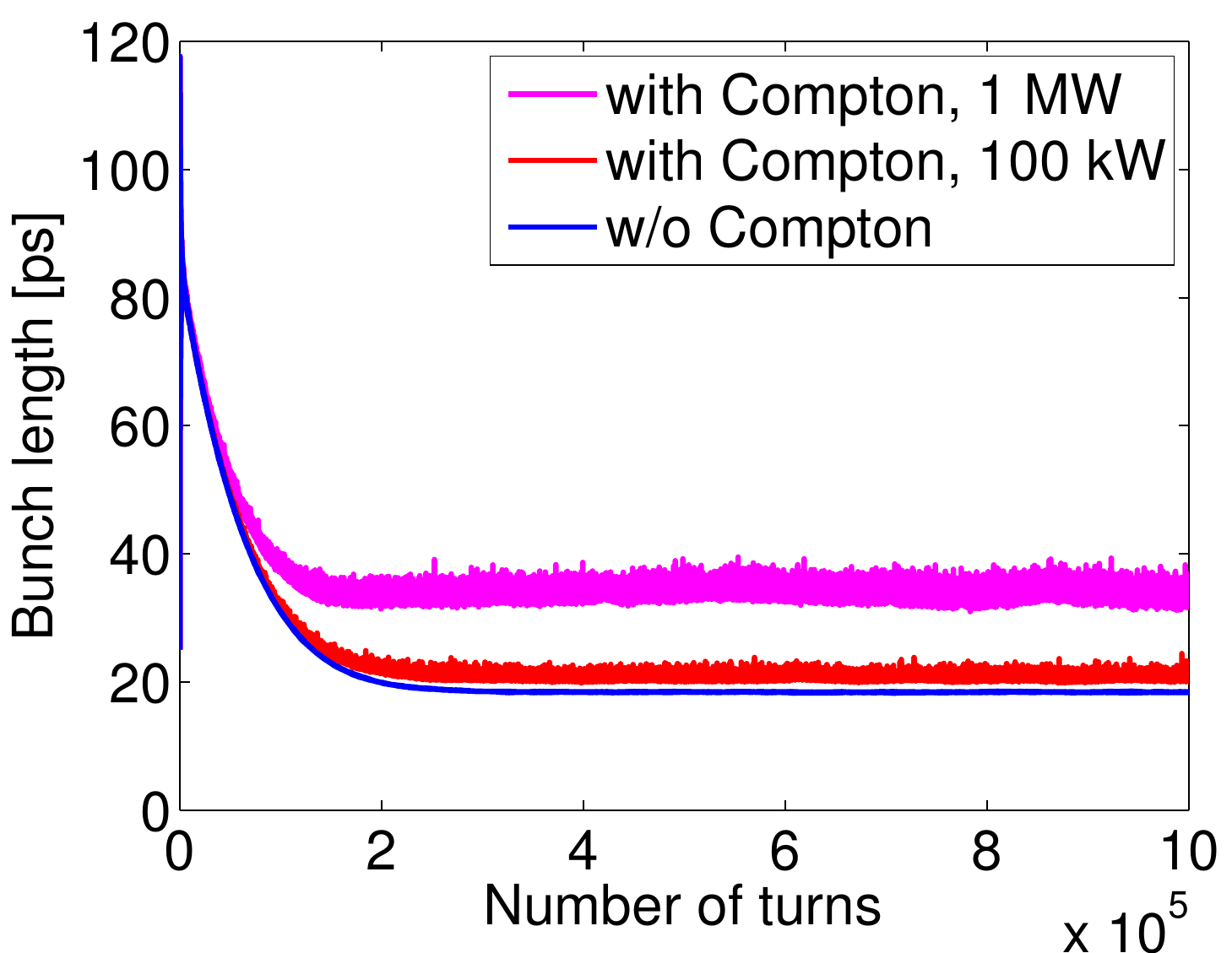}
   \caption{Evolution of the bunch length for different  laser power over approximately 0.5 s. The damping period corresponds to the $\sim2\times10^5$ turns and after the equilibrium is reached.}
   \label{fig:bl}
\end{figure}

\begin{figure}[htb]
   \centering
   \includegraphics*[width=65mm]{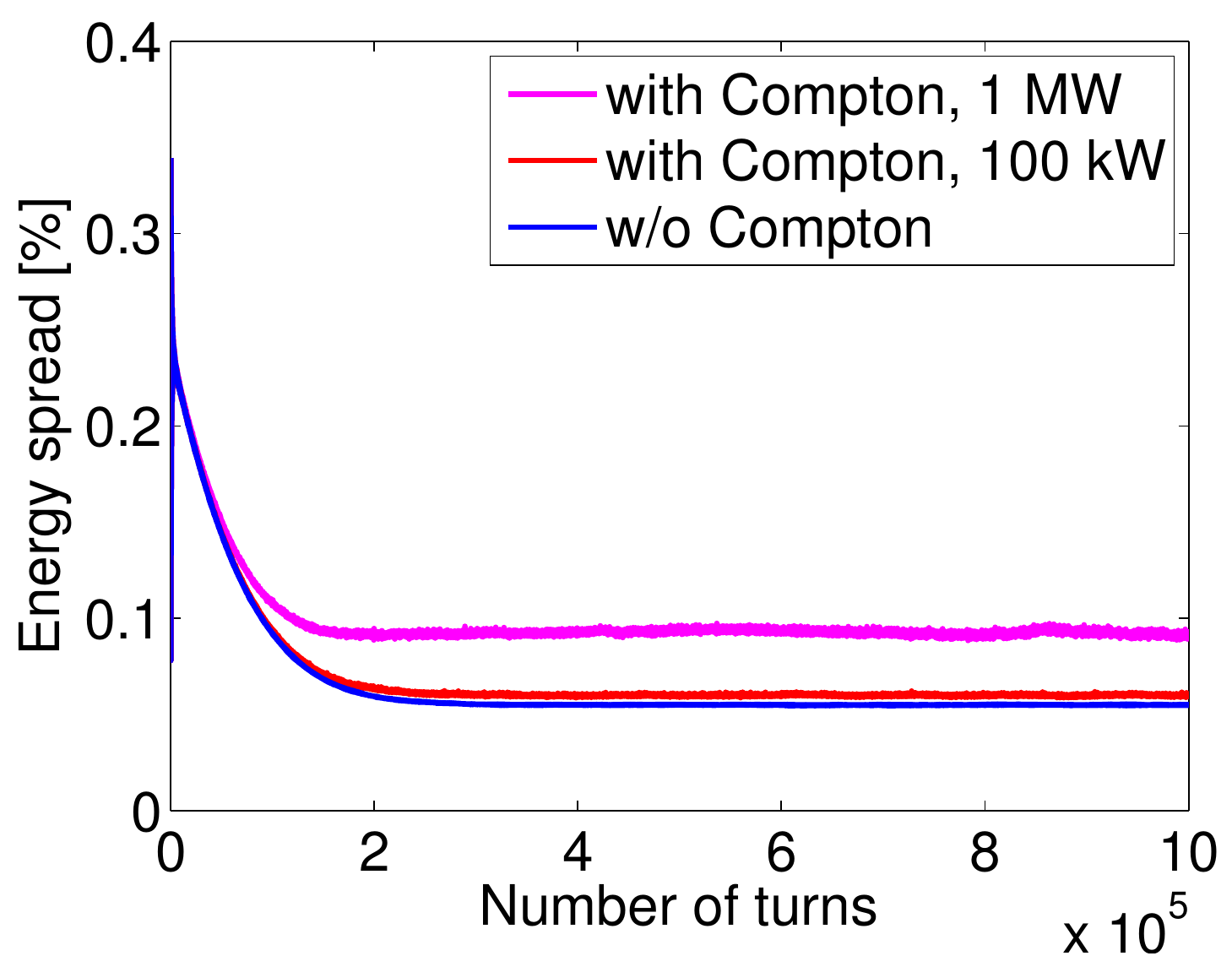}
   \caption{Evolution of the energy spread for different laser power  over approximately 0.5 s.}
   \label{fig:es}
\end{figure}

\begin{figure}[htb]
   \centering
   \includegraphics*[width=65mm]{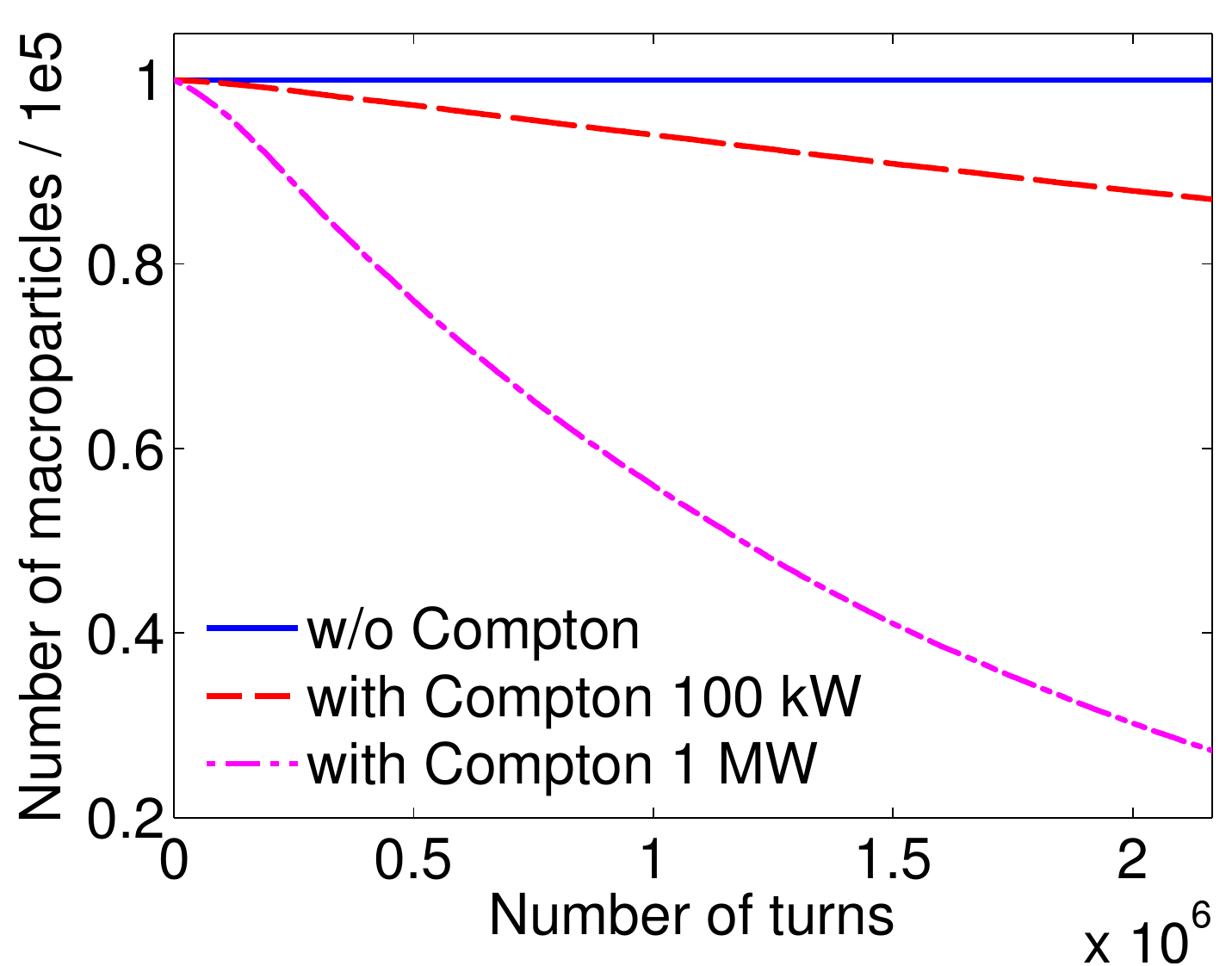}
   \caption{Evolution of the number of macro-particles for different laser power  over 1 s.}
   \label{fig:mp}
\end{figure}

The evolution of the bunch length, the energy spread  and the evolution of the number of macro-particles are presented on Fig.~\ref{fig:bl}, Fig.~\ref{fig:es} and Fig.~\ref{fig:mp} respectively. If the laser power is below 100 kW, the effect of Compton scattering is quite small over  $\sim\!2\times10^6$~turns (1 s) .

For a laser power of 100 kW and 1MW the equilibrium bunch length and the energy spread are respectively $\tau_{e}$ = 22~ps, $\epsilon$ =   0.06~\% and $\tau_{e}$ = 34~ps, $\epsilon$ =   0.1~\%. Fig.~\ref{fig:mp} shows that for 1~MW of laser power, the electrons loss rate is significantly increased.

The main effect of Compton Scattering on the ATF beam is a  reduction of its lifetime. This could be observed by monitoring the evolution of the current in the DR over a second or more.
For the laser power below 100 kW, the Compton effect on the electron beam dynamics at the ATF DR are rather small which makes them difficult to observe. For high enough laser power bunch lengthening could be observed using a streak camera.
\section{Conclusions}
In the context of the Mightylaser project, we simulated the electron beam dynamics under Compton scattering at the ATF DR. Compton scattering reduces the electron beam lifetime in the present set-up but the electrons remaining within the bucket are quickly damped. At higher laser power ($\sim$1 MW) changes in energy spread and bunch length will become observable as well.

\end{document}